\documentclass[12pt,aps,prd,reprint,sort&compress, superscriptaddress, showpacs, nofootinbib, preprintnumbers]{revtex4-1}

\usepackage{graphicx}
\usepackage{amsmath}
\usepackage{amssymb}
\usepackage{dsfont}
\usepackage{amsfonts}
\usepackage[UKenglish]{babel}
\usepackage{hyperref}
\usepackage{nicefrac}
\usepackage{verbatim}
\usepackage{bbm}
\usepackage{color}
\usepackage{multirow}
\usepackage{subcaption}
\usepackage{cleveref}

\begin{document}

\title{$SU(2N_F)$ hidden symmetry of QCD}

\author{L.~Ya.~Glozman}
\email{leonid.glozman@uni-graz.at}
\affiliation{Institut f\"ur Physik, FB Theoretische Physik, Universit\"at Graz, Universit\"atsplatz 5,
8010 Graz, Austria}

\begin{abstract}
Recently a global $SU(4) \supset SU(2)_L \times SU(2)_R \times U(1)_A$ symmetry of the confining Coulombic part of the QCD Hamiltonian  has
been discovered with $N_F=2$. This global symmetry includes both  independent rotations of the
left- and right-handed quarks in the isospin space as well as the chiralspin
rotations that mix the left- and right-handed components of the quark fields.
It has been suggested by  lattice simulations, however, that a
symmetry of mesons  in the light quark sector upon the quasi-zero
mode truncation from the quark propagators is actually higher than $SU(4)$, because
the states from  a singlet and a 15-plet irreducible representations of $SU(4)$ are also degenerate. 
Here we demonstrate that classically QCD, ignoring irrelevant exact zero
mode contributions, has a $SU(2N_F)$ symmetry.
If effects of dynamical chiral symmetry breaking and of anomaly are encoded
in the same near-zero modes, then truncation of these modes should restore
a classical $SU(2N_F)$ symmetry.
Then we show in a Lorentz- and gauge-invariant manner
emergence of a bilocal $SU(4) \times SU(4)$
symmetry in mesons that contains a global $SU(4)$ as a subgroup upon truncation of the quasi-zero modes.
We also demonstrate
that  the confining Coulombic 
part of the QCD Hamiltonian has this bilocal symmetry.
It explains naturally a degeneracy of different irreducible representations of $SU(4)$ observed in lattice simulations.

\end{abstract}

\maketitle

\section{Introduction}

A number of unexpected QCD phenomena has been observed recently in lattice
simulations. A response of mesons with spins $J=0,1$ to an artificial
subtraction of the lowest-lying modes of the overlap Dirac operator from valence quarks has been studied in dynamical $N_F=2$ simulations
\cite{Denissenya:2014poa,Denissenya:2014ywa}. A surprising large degeneracy of mesons, that is larger than the $SU(2)_L \times SU(2)_R \times U(1)_A$ 
symmetry of QCD within the perturbation theory (ignoring  the $U(1)_A$ anomaly) has been discovered.\footnote{For previous lattice
studies on quasi-zero mode extraction see Refs. \cite{LS,GLS}.}
In this paper we always drop a $U(1)_V$ symmetry, that is inessential
for our discussion.
 Apriori one
expects that such a procedure would remove the chiral symmetry breaking
dynamics from hadrons, because the quark condensate of the vacuum is directly
related to a density of the near-zero modes of the Dirac operator via the
Banks-Casher relation \cite{BC}. Obviously, this truncation deforms QCD and
the quark field becomes nonlocal in configuration space. The gluodynamics is kept
intact. Such  truncation is a gauge-invariant procedure
and does not violate
the Lorentz-invariance, because each eigenvalue of the Dirac operator in a given gluonic background is a Lorentz scalar and one does not affect the
Lorentz transformation properties of the quark field. 

One expects that after truncation correlators of operators that are connected with each other via the $SU(2)_L \times SU(2)_R$
transformations would become identical. If hadrons survive this  "surgery", then
masses of chiral partners should  be equal.
It has turned out that a very clean exponential decay of  correlators was observed in all  $J=1$ mesons. This
implies that confined bound  states survive the truncation ~\cite{Denissenya:2014poa}.
  In the $J=0$
sector, while all $J=0$ correlators become identical, the ground
states  disappear from the spectrum, because
there is no exponential decay of the corresponding correlators: The quasi-zero modes  are crucially 
important for the very existence of the (pseudo) Goldstone bosons, which is not surprising.

It has also been found that the truncation restores in hadrons
not only $SU(2)_L \times SU(2)_R$ symmetry, that is broken in QCD dynamically via the quark condensate, but also a $U(1)_A$ symmetry, which is broken  anomalously. From this fact one concludes that the
same lowest-lying modes of the Dirac operator are responsible for both
$SU(2)_L \times SU(2)_R$ and $U(1)_A$ breakings which is consistent with the
instanton-induced mechanism of both breakings \cite{H,S,DP}.

However, not only a degeneracy within
the $SU(2)_L \times SU(2)_R$ and $U(1)_A$ multiplets was detected. A larger
degeneracy that includes all possible chiral multiplets of the $J=1$
mesons was observed, which was completely unexpected. This larger degeneracy implies a symmetry that is larger than 
$SU(2)_L \times SU(2)_R \times U(1)_A$.

A symmetry group that drives this degeneracy has been reconstructed in 
Ref.~\cite{Glozman:2014mka}, that is $SU(4) \supset SU(2)_L \times SU(2)_R \times U(1)_A$. Transformations of this group "rotate" the fundamental vector
$(u_L, u_R, d_L, d_R)^T$ and include both  independent rotations of
the left- and right-handed quarks in the isospin space as well as rotations
in the chiralspin space that mix the left- and right-handed components of the quark fields. 
This symmetry implies that there are no magnetic
interactions between quarks and a meson represents a dynamical quark-antiquark
system connected by a confining electric field. Such a system was interpreted
as a dynamical QCD string and the $SU(4)$ symmetry was identified to be a symmetry of confinement.

The $SU(4)$-transformation properties of the $\bar q q$ operators have been
studied in Ref.  \cite{Glozman:2015qva}. It has also been found  where
this symmetry is hidden - the $\gamma_0$-part of the quark-gluon interaction term is a
$SU(4)$-singlet. This part of the Lagrangian is responsible for  
interaction of quarks
with the chromo-electric field. Interactions of quarks with the
chromo-magnetic field are not $SU(4)$-invariant and transform as a 15-plet
of $SU(4)$.  It is  a generic
property of any local gauge-invariant theory. With  $N_F$ light flavors the
symmetry group is obviously $SU(2N_F)$.

The role of this symmetry in hadrons can be   seen using the Hamilton
language \cite{Glozman:2015qva}. In Coulomb gauge the QCD Hamiltonian  \cite{CL} consists
of a few terms:  a gauge field dynamics,
a quark field term, an interaction  between the quark field and the
chromo-magnetic  field, and the "Coulombic" part that represents
an instantaneous  chromo-electric interaction between the color-charge densities
located at different spatial points. The color-charge density operator includes both the charge density of the gluonic field, that is obviously independent from quark isospin and its chirality, as well as the quark charge density.
Since the latter  is a $SU(4)$-singlet, the whole "Coulombic" part
of the QCD Hamiltonian is also a $SU(4)$-singlet. Note that while the Hamilton
description in Coulomb gauge is not a covariant description,  the observable color-singlet gauge-invariant quantities are Lorentz-invariant. The Hamiltonian provides a
very convenient description of a system in its rest-frame. 

With the static color sources
the Coulombic part of the Hamiltonian implies a confining  linear potential \cite{Z}.
The  $SU(4)$ symmetry can be viewed as a symmetry
of a confining dynamical QCD string in the light quark sector.

There is another implication from this. 
The interaction of quarks with the chromo-magnetic field, that explicitly
breaks the $SU(4)$ symmetry,  is located in the confinement regime
only in the near-zero mode zone and is responsible
for  all symmetries   $SU(2)_L \times SU(2)_R$,  $U(1)_A$ and $SU(4)$ breakings. A truncation
of the near-zero modes filters out a confining dynamics in the system.

Meanwhile emergence of  $SU(4)$ has been confirmed in the $J=2$ meson
sector \cite{Denissenya:2015mqa} and in baryons
\cite{Denissenya:2015woa}.

The global $SU(4)$ symmetry cannot explain, however, a degeneracy of the
$J=1$  states $\rho, \rho', \omega, \omega', a_1, h_1, b_1$, that form a 
$SU(4)$ 15-plet, and of a singlet $f_1$.\footnote{The latter degeneracy requires,
however, a confirmation, because an effective mass plateau for the $f_1$ state
is not convincing \cite{Denissenya:2014ywa}.} It has been suggested in Ref.~ \cite{Glozman:2015qva} that this degeneracy, if confirmed, should imply existence of a larger symmetry that includes $SU(4)$ as a subgroup.
 Cohen has found very recently that no higher symmetry  is phenomenologically
acceptable,  that
would connect {\it local} quark bilinears from the 15-plet and singlet of $SU(4)$
within  one and the same irreducible representation of some higher group \cite{TDC}.

Here we show that classically QCD  has, excluding irrelevant
exact zero mode contributions, a $SU(2N_F)$ symmetry.
Chiral symmetry spontaneous breaking is encoded in the near-zero
modes of the Dirac operator, as it follows from the Banks-Casher
relation. If effects of anomaly are also encoded in the same near-zero
modes, then truncation of the near-zero modes should restore a classical
$SU(2N_F)$ symmetry.
 Given this symmetry
we demonstrate 
in a gauge- and Lorentz-invariant manner emergence at $N_F=2$ of a
{\it bilocal} $SU(4) \times SU(4)$ symmetry in mesons and of a trilocal  
$SU(4) \times SU(4) \times SU(4)$ symmetry in baryons upon elimination of
the quasi-zero modes of the Dirac operator. We also show
that  the confining part of the QCD Hamiltonian is actually
invariant not only under global $SU(4)$ transformations, but is also a singlet
under bilocal $SU(4) \times SU(4)$ transformations.  An irreducible dim=16 representation of $SU(4) \times SU(4)$ combines both the $SU(4)$-singlet and the 15-plet into one
irreducible representation, which explains a degeneracy of $f_1$ with the 
15-plet mesons. This symmetry implies invariance upon independent instantaneous
$SU(4)$ transformations
of the quark fields at  different space points $\boldsymbol{x}$ and $\boldsymbol{y} $ and is intrinsically nonlocal and 
 cannot be represented by local composite operators.

\section{ Global and space-local  symmetries of the Coulombic
interaction for massless quarks}

Consider the quark-gluon interaction part of the QCD Lagrangian with
$N_F$ massless flavors in Minkowski space-time:

\begin{equation}
\label{Lagrangian}
 \overline{\Psi} (\mathds{1}_{\textsc{F}} \otimes i \gamma^{\mu} D_{\mu}) \Psi = \overline{\Psi} (\mathds{1}_{\textsc{F}} \otimes i \gamma^0 D_0)  \Psi 
  + \overline{\Psi} (\mathds{1}_{\textsc{F}} \otimes i \gamma^i D_i)  \Psi\; . 
\end{equation}

\noindent
The first term describes an interaction of the quark charge density 
$\rho(x) = \bar \Psi (x) \gamma^0 \Psi(x)$ with the chromo-electric  
part of the gluonic field. The second term contains an
interaction of the spatial current density with the chromo-magnetic field.

The chromo-electric part of the interaction Lagrangian is invariant under a global and space-local ($\boldsymbol{x}$ - dependent) $SU(2)_{CS} \supset U(1)_A$ and
$SU(2N_F)$ transformations of the quark field. 
The $SU(2)_{CS} \supset U(1)_A$  transformations are
defined as

\begin{align}
\label{V-def}
  \Psi \rightarrow  \Psi^{'} &= e^{i (\mathds{1}_{\textsc{F}} \otimes \frac {\boldsymbol{\varepsilon} \cdot \boldsymbol{\Sigma}}{2})} \Psi  \; .
\end{align}

\noindent
with the following generators 
\begin{align}
\label{sigma}
\boldsymbol{\Sigma} = \{ \gamma^0, i \gamma^5 \gamma^0, -\gamma^5 \} \; ,  
\end{align}

\noindent
that form an $SU(2)$ algebra
\begin{align}
 [\Sigma^i,\Sigma^j] = 2 i \epsilon^{i j k} \, \Sigma^k \; .
\end{align}

\noindent
An imaginary three-dimensional space in which these rotations are
performed is refered to as the {\it chiralspin} (CS) space \cite{Glozman:2014mka,Glozman:2015qva}. Upon the  chiralspin rotations
the right- and left-handed components of the fermion fields get mixed.
It is similar to the well familiar
concept of the isospin space: The electric charges of particles are conserved
quantities, but rotations in the isospin space mix particles with different
electric charges.

If the rotation vector $\boldsymbol{\varepsilon}$ is space-time independent,
then this transformation is global. If the rotation parameters are different at different 
space points $\boldsymbol{x}$,   $\boldsymbol{\varepsilon}(\boldsymbol{x})$,
then we refer to such a rotation as a space-local transformation.
The first term in eq. (\ref{Lagrangian}) is invariant with
respect to both global and space-local $SU(2)_{CS}$ transformations.

When we combine the $SU(2)_{CS}$ rotations with the chiral $SU(N_F)_L \times SU(N_F)_R$ transformations into one larger group, then we arrive at a
$SU(2N_F)$ group. For example, in  case of two flavors the $SU(4)$ transformations

\begin{align}
\label{W-def}
\Psi \rightarrow  \Psi^{'} &= e^{i \boldsymbol{\epsilon} \cdot \boldsymbol{T}/2} \Psi\; ,
\end{align}
are defined through the following set of 15 generators:
\begin{align}
 \{(\tau^a \otimes \mathds{1}_D), (\mathds{1}_F \otimes \Sigma^i), (\tau^a \otimes \Sigma^i) \} \;. 
\end{align}

\noindent
With the space-time independent $(2N_F)^2 -1$-dimensional  rotation vector
$\boldsymbol{\epsilon}$ the corresponding symmetry is global, while with
the space-dependent rotation $\boldsymbol{\epsilon}( \boldsymbol{x})$ it is space-local.

The magnetic part of the interaction Lagrangian does not admit this higher
symmetry and is invariant only with respect to global $SU(N_F)_L \times SU(N_F)_R \times U(1)_A$ chiral transformations.

The  higher symmetry of the electric part of the
interaction Lagrangian is generic for any local gauge-invariant theory and
has significant implications for confinement in QCD.

\section{Zero modes of the Dirac operator and symmetries of  Euclidean QCD}

Enlarged symmetry, reviewed in the Introduction, is obtained
in lattice simulations upon subtraction of the near-zero modes of the Dirac
operator. This means that this symmetry should be encoded in the Euclidean QCD once the near zero modes are removed.
In this section we discuss symmetry properies of the nonperturbatively defined
QCD in Euclidean space-time and demonstrate that indeed once the near zero
modes, that are responsible for both spontaneous and anomalous chiral symmetries breakings, are subtracted the QCD partition function is $SU(2N_F)$
symmetric. As an introduction to the Euclidean field theory
we recommend the textbook \cite{Ramond}.

The Lagrangian in Euclidean space-time with $N_F$ degenerate massive quarks
in a given gauge configuration is:

\begin{equation}
{\cal {L}} = \Psi^\dag(x)( \gamma_\mu D_\mu + m) \Psi(x),
\label{lag}
\end{equation}

\noindent
with

\begin{equation}
D_\mu = \partial_\mu + i g\frac{t^a}{2} A^a_\mu,
\end{equation}

\noindent
where $A^a_\mu$ is the gluon field configuration and
$t^a$ are the $SU(3)$-color generators.

In Euclidean space the Grassmannian fields $\Psi(x)$ and $\Psi^\dagger(x)$
are completely independent from each other. Different parts of
the Lagrangian (\ref{lag}) have different symmetries. For example,
the mass term $\Psi^\dag(x)\Psi(x)$ is invariant under a $U(1)_A$ 
transformation

\begin{equation}
\Psi(x) \rightarrow e^{i\alpha \gamma_5}\Psi(x); ~~~~
\Psi^\dagger(x) \rightarrow \Psi^\dagger(x)e^{-i\alpha \gamma_5}.
\end{equation}

\noindent
At the same time the kinetic term, $\Psi^\dag(x)( \gamma_\mu D_\mu)\Psi(x)$,
breaks this symmetry, because the $\gamma_5$ matrix anticommutes with
all $\gamma_\mu$ matrices, $\gamma_5\gamma_\mu + \gamma_\mu \gamma_5 = 0$.
The same is true with respect to the axial part of the 
$ SU(N_F)_L \times SU(N_F)_R$ transformation.\footnote{This definition
of the chiral transformation in Euclidean space is consistent with the 
Lorentz ($SO(4)$)- transformation properties of the $\Psi^\dag$ field, see
for details ref. \cite{Ramond}, and is used in the literature \cite{We}.
Very often another definition of the chiral transformation
is given,
 
 $$ {L} = \bar \Psi(x)( \gamma_\mu D_\mu + m) \Psi(x),$$

$$\Psi(x) \rightarrow e^{i\alpha \gamma_5}\Psi(x); ~~~~
\bar \Psi(x) \rightarrow \bar \Psi(x) e^{i\alpha \gamma_5},$$ 
\noindent
which is
inconsistent, however, with the $SO(4)$-rotation  properties
of the field $\bar \Psi(x)$, that transforms as $\Psi^\dag(x)$. When
calculating  the fermion determinant the field $\bar \Psi(x)$ is implicitly
substituted through  $\Psi^\dag(x)$, so the fermion
determinant and generating functional have correct Lorentz transformation
properties, see for a transparent exposition
ref. \cite{Co}. Consequently all real Euclidean lattice results are correct
since they do not depend on a semantical issue what part of the Lagrangian
above should be called chirally symmetric and what part - chiral symmetry breaking.}

What are symmetry properties of both parts under the $SU(2)_{CS}$ and 
$SU(2N_F)$ transformations?
The Euclidean $\gamma_4$ coincides with the Minkowski
$\gamma_0$ and  $\gamma_5$ matrices in both spaces are equal.
Then we can define Euclidean $SU(2)_{CS}$  transformations
through  generators that satisfy a $SU(2)$ algebra,

\begin{align}
\label{sigma}
\boldsymbol{\Sigma} = \{ \gamma^4, i \gamma^5 \gamma^4, -\gamma^5 \} \; .  
\end{align}

\noindent
Combining the $SU(2)_{CS}$ generators with the $SU(N_F)$ flavor generators
into a larger algebra like in eq. (6) we arrive at the Euclidean $SU(2N_F)$ transformations.

It is then obvious that 
the kinetic term in (\ref{lag}) breaks both symmetries. It is invariant
only under flavor transformation $SU(N_F)$. It is important to
understand the underlying reason why these symmetries are missing in
the kinetic term. The reason is that these symmetries are absent for a
quark that is "on-shell". The "on-shell" quark is described by the
Dirac equation,

\begin{equation}
 \gamma_\mu D_\mu  \Psi_0(x) = 0.
\label{dir}
\end{equation}

\noindent
Its solution is traditionally called a zero mode.
Zero modes are solutions of the Dirac equation with the gauge
configurations of a nonzero topological charge. They are absent
in  gauge configurations with $Q=0$. The difference of numbers of
the left-handed
and right-handed zero mode solutions is according to the Atiyah-Singer
theorem  fixed by the topological charge $Q$ of the gauge configuration:

\begin{equation}
 n_L - n_R = Q.
\label{AZ}
\end{equation}

\noindent
For example, with a gauge configuration of $Q=1$ there is only a left-handed
zero mode and there is no right-handed zero mode solution \cite{i1,i2}. Some $SU(2)_{CS}$ transformations rotate
the right-handed solution into the left-handed solution and vice versa.
Consequently,  the zero mode  explicitly violates the $SU(2)_{CS}$ and
$SU(2N_F)$ symmetries. Similar analysis can be done for any $Q \neq 0$.
In other words, the zero modes  intrinsically introduce an asymmetry between
the left- and right-handed degrees of freedom and manifestly break the 
$SU(2)_{CS}$. The latter invariance  is possible only if there is no asymmetry
between the left and the right.

Exact zero modes are absent in the $Q=0$ sector of Euclidean QCD. Consequently, in this sector
the symmetry properties of the kinetic term could be quite different.

The fermionic part of the QCD partition function in a volume $V$
in the $Q=0$ sector is given as

\begin{equation}
Z^V_{Q=0} = \int D\Psi D\Psi^\dagger 
e^{\int d^4x \Psi^\dag(x)( i\gamma_\mu D_\mu + im) \Psi(x)}. 
\label{Z0}
\end{equation}

\noindent
The Grassmannian fields $\Psi(x)$ and $\Psi^\dagger(x)$ are
 defined in the following standard way.
 The hermitian Dirac operator, $i \gamma_\mu D_\mu$,  in a given gluonic background has in a finite volume $V$ a discrete spectrum with real eigenvalues $\lambda_n$:

\begin{equation}
i \gamma_\mu D_\mu  \Psi_n(x) = \lambda_n \Psi_n(x).
\label{ev}
\end{equation}

\noindent
The nonzero eigenvalues come in pairs $\pm \lambda_n$,
because 

\begin{equation}
i \gamma_\mu D_\mu  \gamma_5 \Psi_n(x) = -\lambda_n \gamma_5 \Psi_n(x).
\end{equation}

\noindent
We can expand fields $\Psi(x)$ and $\Psi^\dagger(x)$  over a complete
and orthonormal set $\Psi_n(x)$:

\begin{equation}
 \Psi(x) = \sum_{n} c_n \Psi_n(x), ~~~~~ 
 \Psi^\dagger(x) = \sum_{k}\bar {c}_k \Psi^\dag_k(x),
\end{equation}
 where $\bar {c}_k,c_n$ are Grassmannian numbers.
 For all eigenvectors with the nonzero
eigenvalues $\lambda_n \neq 0$ we can replace

\begin{equation}
\gamma_\mu D_\mu   \Psi_n(x) \rightarrow -i \lambda_n  \Psi_n(x).
\label{sub}
\end{equation}

\noindent
This substitution effectively eliminates the $\gamma_\mu D_\mu$ operator
and replaces it with the Lorentz-scalar $\lambda_n$.
Then the partition function (\ref{Z0}) takes the following form

\begin{equation}
Z^V_{Q=0} = \int \prod_{k,n} d\bar {c}_k dc_n  
e^{\sum_{k,n}\int d^4x 
 \bar {c}_k  c_n (\lambda_n + im) \Psi_k^\dag(x) \Psi_n(x)}. 
\label{ZZ}
\end{equation}
 
 \noindent
Now we can directly read-off symmetry properties of this
partition function. Since this functional contains only a superposition of
terms  $\Psi_k^\dag(x) \Psi_n(x)$ it is precisely $SU(2)_{CS}$ and
$SU(2N_F)$ symmetric, because

\begin{equation}
(U\Psi_k(x))^\dag U \Psi_n(x) = \Psi^\dag_k(x) \Psi_n(x),
\label{tt}
\end{equation}
 
\noindent
where $U$ is any unitary transformation from the
groups $SU(2)_{CS}$ and
$SU(2N_F)$ , $ U^\dag = U^{-1}$. 
The exact zero modes, for which the equation (\ref{tt}) does not hold,
 do
not contribute to this partition function.
We conclude that classically the Euclidean QCD in a finite volume $V$
without the exact zero modes, i.e. in the Q=0 sector,  is invariant with respect to both
global and space-local  $SU(2N_F)$ transformations.

Quantization implies an integration over fields  $\Psi(x)$ and $\Psi^\dagger(x)$. In this case the $U(1)_A$ symmetry is broken anomalously,
because the measure $D\Psi D\Psi^\dagger$ is not invariant upon a local
$U(1)_A$ transformation \cite{FU}. Since the $U(1)_A$ is a subgroup of  
$SU(2)_{CS}$, the anomaly also breaks the $SU(2)_{CS}$ symmetry.

In the thermodynamic limit $V \rightarrow \infty$ the otherwise finite
lowest eigenvalues $\lambda$ condense around zero, $\lambda \rightarrow 0$,
and according to  the Banks-Casher relation,

\begin{equation}
\lim_{m \rightarrow 0} <0|\bar \Psi(x) \Psi(x)|0> = -\pi \rho(0)~,
\end{equation}

\noindent
provide a nonvanishing quark condensate in Minkowski space. Here a sequence of limits
is important. First an infinite volume limit must be taken and
only then a chiral limit. The quark condensate in Minkowski space-time
breaks all $SU(2)_{CS}$, $SU(N_F)_L \times SU(N_F)_R$ and
$SU(2N_F)$ symmetries to the vector flavor symmetry $SU(N_F)_V$.
Consequently the new $SU(2)_{CS}$ and $SU(2N_F)$ symmetries are
broken both by the condensate and anomalously.

Effect of dynamical chiral symmetry breaking is encoded in the near-zero 
modes of the Dirac operator. If effects of anomaly are also encoded in
the near-zero modes, as suggested e.g. by the instanton mechanism of
both breakings \cite{SS}, then
removal on lattice
of  eigenmodes with lowest $\lambda$
should restore consequently not only chiral $SU(N_F)_L \times SU(N_F)_R$  and $U(1)_A$ symmetries, but also
a larger $SU(2N_F)$ symmetry, in agreement with observations reviewed in the
Introduction.

In the thermodynamic limit, $V \rightarrow \infty$, the $Q=0$
partition function coincides, up to an inessential normalization
factor, with the total QCD partition function
taken at zero theta-angle, $\theta =0$ \cite{LeuS}. It approaches 
the full QCD partition function rather fast, as $1/V$ \cite{N}.

What we have described above is a symmetry of QCD in the Q=0 sector.
What about symmetries in all other sectors? They  all contribute
to the QCD partition function  at $\theta = 0$. We can work in any of them 
\cite{LeuS,N}. In this case the partition function will contain the
zero mode contributions. For example, in the case Q=1 there will appear 
terms that contain a zero mode eigenfunction: $\Psi^\dagger_0(x) \Psi_0(x)$,
$\Psi^\dagger_k(x) \Psi_0(x)$ and $\Psi^\dagger_0(x) \Psi_n(x)$. They all
explicitly break both $SU(2)_{CS}$ and $SU(2N_F)$ symmetries, because
in this case the relation (19) does not hold. 
 However, these zero mode contributions
are completely irrelevant  in the Green functions
and observables in the thermodynamic limit since they vanish as $1/V$ \cite{LeuS,A}.
Consequently, in the limit $V \rightarrow \infty$ we will approach the
same result in any sector.

Summarizing, if effects of anomalous and dynamical chiral symmetry
breakings are encoded in the same near-zero modes, then removal of these modes should restore a hidden classical $SU(2N_F)$ symmetry.

We have discussed in this section  a symmetry of QCD defined
nonperturbatively. With the noninteracting fermions or within the
perturbation theory there is no  $SU(2)_{CS}$ symmetry,
because for on-shell massless 	quarks chirality is a conserved
quantum number. Any Feynman diagram can be continued from Minkowski
to Euclidean space via the Wick rotation. Consequently, a symmetry
of QCD within the perturbation theory ignoring the $U(1)_A$ anomaly is only $SU(N_F)_L \times SU(N_F)_R \times U(1)_A$.

\section{$SU(2N_F) \times  SU(2N_F)$ and $SU(2N_F) \times  SU(2N_F) \times  SU(2N_F)$ emergent symmetries in mesons and baryons}

We have considered above emergence of 
 global and space-local $SU(2N_F)$ symmetries of a quark in a given
gauge background. What symmetries should we expect in hadrons upon
the quasi-zero modes elimination?

Hadron spectra are obtained from the correlation functions calculated
with the gauge-invariant source operator. At each time slice "t" a meson
correlator contains minimum the lowest Fock component consisting
of  valence quark and antiquark  located at different space points 
$\boldsymbol{x}$ and $\boldsymbol{y}$. Both quark and antiquark interact with the same gauge configuration according to eq. (\ref{lag}). Consequently, all
arguments of the previous section apply independently for both quark and antiquark. Since the $SU(2N_F)$ invariance is space-local, we can perform independent $SU(2N_F)$ rotations at points $\boldsymbol{x}$ and $\boldsymbol{y}$ with different rotation parameters. One then concludes that
the meson correlation function with the quark-antiquark valence content
has a {\it bilocal} $SU(2N_F) \times SU(2N_F)$ symmetry. A symmetry of 
higher Fock components is larger, but the whole correlator 
has a symmetry of the lowest quark-antiquark component. Obviously,
averaging over gauge configurations will not change  this symmetry property.

The same arguments apply to baryons and in this case we can expect a trilocal
$SU(2N_F) \times SU(2N_F) \times SU(2N_F)$ symmetry. 

\section{ $SU(2N_F) \times SU(2N_F)$ bilocal symmetry of confinement
in Coulomb gauge}

In previous sections we discussed a Lorentz- and gauge-invariant derivation
of emergence of bilocal and trilocal symmetries of hadrons upon the quasi-zero
mode elimination. A clear connection with  confinement physics
 is missing in those derivations, however. To get an insight
consider the QCD Hamiltonian in Coulomb gauge in Minkowski space \cite{CL}:

\begin{equation}
H_{QCD} = H_E + H_B
\nonumber
\end{equation}

 \begin{equation}
 + \int d^3 x \Psi^\dag({\boldsymbol{x}}) 
[-i \boldsymbol{ \alpha} \cdot \boldsymbol{\nabla} + \beta m ]  \Psi(\boldsymbol{x})
+ H_T + H_C,
\end{equation}

\noindent
where the transverse (magnetic) and Coulombic interactions are:

\begin{equation}
H_T = -g \int d^3 x \, \Psi^\dag({\boldsymbol{x}}) \boldsymbol{\alpha} 
\cdot \frac{t^a}{2} \boldsymbol{A}^a(\boldsymbol{x}) \, \Psi(\boldsymbol{x}) \; , 
\end{equation}

\begin{equation} 
H_C = \frac{g^2}{2} \int  d^3 x \, d^3 y\, J^{-1} \ \rho^a(\boldsymbol{x})  F^{ab}(\boldsymbol{x},\boldsymbol{y}) \, J \, \rho^b(\bf y) \; ,
\label{c}
\end{equation}
 
\noindent
with $J$ being  Faddeev-Popov determinant, $\rho^a(\boldsymbol{x})$
is a color-charge density and $F^{ab}(\boldsymbol{x},\boldsymbol{y})$ is a confining
Coulombic kernel.

The fermionic and transverse (magnetic) parts of the Hamiltonian
have at $m \rightarrow 0$ $SU(N_F)_L \times SU(N_F)_R$ and $U(1)_A$ global chiral symmetries.
A symmetry of the confining Coulombic part is higher, however.
It is not only invariant under  $SU(N_F)_L \times SU(N_F)_R$ and $U(1)_A$ global chiral transformations, but is also a singlet
with respect to  bilocal 
$SU(2N_F) \times SU(2N_F)$ transformations. Indeed, the color charge
density $\rho^a(\boldsymbol{x})$ at a space point $\boldsymbol{x}$ is a singlet with respect
to the  $SU(2N_F)$ rotations. The same is true for the
color charge density at a space point $\boldsymbol{y}$. However, the
rotations of the fermion field at the space points  $\boldsymbol{x}$ and 
$\boldsymbol{y}$ can be completely independent from each other, with different
rotation parameters. Consequently, the Coulombic part of the QCD Hamiltonian
is not only a singlet with respect to global $SU(2N_F)$
transformations, but is also invariant under independent  $SU(2N_F)$ transformations of the fermion field at points $\boldsymbol{x}$
and $\boldsymbol{y}$. 

We conclude that the Coulombic confining part of the QCD Hamiltonian has a
bilocal  $SU(2N_F) \times SU(2N_F)$
symmetry.

\section{Implications}

 An
irreducible representation of the $SU(4) \times SU(4)$ group is 16-dimensional
and is a direct sum of the 15-plet and singlet of $SU(4)$. Consequently, a direct prediction of this symmetry is a degeneracy of the $SU(4)$-singlet and
of the $SU(4)$ 15-plet, in agreement with the lattice observations reviewed in the
Introduction. Since this symmetry is bilocal it cannot be represented
by  local composite operators and our result is consistent with  
conclusions of Ref. \cite{TDC}. 

A $N_F=2$ model with the  Hamiltonian (\ref{c}) has been solved in the past
in Ref. {\cite{WG}}. In that work a confining linear instantaneous potential has been assumed as
a confining kernel $F^{ab}(\boldsymbol{x,y})$ and a large $N_c$ meson
spectrum has been obtained upon solution of the gap and Bethe-Salpeter equations. At high meson spins, where  effects of  chiral symmetry
breaking in the vacuum became irrelevant,   the spectrum
was observed to be highly degenerate and all parity-chiral multiplets of states
of the same spin, namely $(0,0), (0,0), (1/2,1/2)_a, (1/2,1/2)_b, (0,1)+(1,0)$
had the same mass. This result was obtained both numerically and analytically. 
This degeneracy is nothing else but a  $SU(4) \times SU(4)$
symmetry discussed above, because a dim=16 irreducible representation
of $SU(4) \times SU(4)$ is a direct sum $(0,0) + (0,0) + (1/2,1/2)_a + (1/2,1/2)_b  +  (0,1)+(1,0)$ of  irreducible representations of the parity-chiral group.
This result does not necessarily mean that confinement
in QCD in  the light quark sector is reduced to an instantaneous linear potential, but it does illustrate a generic symmetry property and implications of the Coulombic
Hamiltonian. Namely, when the chiral symmetry breaking dynamics is switched 
off a spectrum reveals a  bilocal $SU(4) \times SU(4)$ symmetry of the Hamiltonian.

\section{Interpolation between the heavy quark and chiral limits}

In the heavy quark limit \cite{IW} a matrix element of the charge density
operator is diagonal in flavor and spin spaces and the Coulombic Hamiltonian has a nonrelativistic
 $SU(2N_F)_{SF} \times SU(2N_F)_{SF}$ spin-flavor symmetry.
A spectrum is $SU(2N_F)_{SF} \times SU(2N_F)_{SF}$ symmetric. It is a symmetry of the nonrelativistic quark model. 

 In the chiral limit the charge density operator is diagonal in flavor
and chirality spaces and the Coulomb Hamiltonian has a $SU(2N_F) \times SU(2N_F)$
symmetry discussed in previous sections.
 The same Coulomb Hamiltonian
 has in the opposite limits of QCD two different symmetries.

\section{Conclusions}

We have demonstrated the following  points.

1. Classically QCD has, excluding  irrelevant exact zero mode
contributions,  $SU(2)_{CS}$ and $SU(2N_F)$ symmetries. Since these symmetries
are not present at the level of the QCD Lagrangian and are visible only when
we subtract exact zero modes, we call them  as  hidden symmetries of QCD. 
These symmetries are broken in the real world
by the axial anomaly and by the condensate.

2. Truncation of the near-zero modes that encode both dynamical
chiral symmetry breaking and anomaly restores  hidden classical symmetries of QCD. 

3. In a Lorentz- and gauge-invariant manner we have shown
that elimination of the near-zero modes of quarks leads
to  $SU(2N_F) \times SU(2N_F)$ and 
$SU(2N_F) \times SU(2N_F) \times SU(2N_F)$
symmetries in hadrons.

4. The confining Coulombic part of the QCD Hamiltonian in Coulomb gauge
has a bilocal $SU(2N_F) \times SU(2N_F)$ symmetry. This  symmetry
implies a $SU(2N_F) \times SU(2N_F)$-symmetric spectrum in mesons 
and $SU(2N_F) \times SU(2N_F) \times SU(2N_F)$ symmetry
in baryons
in case when 
other parts of the Hamiltonian that break this symmetry become inessential.
This property has been illustrated with a solvable relativistic  confining
model.

5. This bilocal symmetry of confinement is consistent with 
degeneracy of the $SU(4)$ singlet $f_1$ correlator with the $SU(4)$
15-plet $\rho, \rho', \omega, \omega', h_1, a_1, b_1$  correlators
observed on the lattice after subtraction of the quasi-zero modes of the Dirac operator.

\begin{acknowledgements}
The author thanks Tom Cohen and Christian Lang for numerous  discussions.
On my question what nonrelativistic  symmetry, that
is higher than the spin-isospin $SU(4)_{SI}$, is responsible for a degeneracy
of the $SU(4)_{SI}$-singlet ($I=0,^1S_0$) with the  $SU(4)_{SI}$-15-plet
($I=0,^3S_1$; $I=1,^1S_0$; $I=1,^3S_1$), Tom has pointed out  that
this symmetry is  $SU(4)_{SI} \times SU(4)_{SI}$. This remark was an incentive
for this study.

We acknowledge support from the Austrian Science Fund (FWF)
through the grant P26627-N27. 
\end{acknowledgements}

\end{document}